\documentclass[a4paper,twoside]{article}
\usepackage{amssymb}
\usepackage{amsmath}
\usepackage{amsfonts}
\usepackage{graphicx}

\begin{document}

\title{\Large\bf{
Gravitational collapse in loop quantum gravity 
}}

\author{\\ Leonardo Modesto
 \\[1mm]
\small{ Department of Physics, Bologna University V. Irnerio 46, I-40126 Bologna \& INFN Bologna, EU}
   }

\date{\ } 
\maketitle

\begin{abstract}
In this paper we study the gravitational collapse in loop quantum gravity.
We consider the space-time region inside the Schwarzschild black hole event horizon
and we divide this region in two parts, the first one  where the matter (dust matter) 
is localized and the other (outside) where the metric is Kantowski-Sachs type. 
We calculate the state solving Hamiltonian constraint and we obtain 
a set of three difference equations that give a regular and natural evolution beyond 
the classical singularity point in  ``$r=0$" localized.  

\end{abstract}

\section*{Introduction}
Quantum gravity, the theory that wants reconcile general relativity and quantum
mechanics, is one of  major problem in theoretical physics today.
The lesson of general relativity is that also the space-time is dynamical, 
then it is not possible to study the other interaction on a fixed background.
The background itself is a dynamical field.
 
 One of more diffuse theory of quantum gravity is the theory called
of  ``loop quantum gravity" \cite{book}. This is one of the non perturbative and 
background independent approach to quantum gravity
(another non perturbative approach to quantum gravity is called "asymptotic safety 
quantum gravity" \cite{R}). 

In this paper we will apply ideas suggested by full loop quantum gravity 
to a minisuperspace model where we will impose symmetries on the full
metric to obtain a reduced model. It is possible to implement 
the Dirac quantization program following the fundamental ideas 
of loop quantum gravity.

Some interesting results in this theory 
are related to the problem of space-like singularity. In fact it was
shown in \cite{Boj}, \cite{MAT}, \cite{work1} and \cite{work4} that it is 
possible to solve the cosmological singularity problem and the
black hole singularity problem by using the ideas developed in full loop
quantum gravity theory. 
In \cite{BR} the problem of black hole singularity has been 
analyzed also under the contest of "asymptotic safety 
quantum gravity"; in that paper authors showed that non perturbative 
quantum gravity effects give a much less singular Schwarzschild metric.

In this paper we would like to study the gravitational collapse of 
a dust sphere inside the event horizon \cite{NS}. 
We consider the phase of the gravitational collapse when all  (dust)  matter  
has crossed the Schwarzschild radius. In this phase the symmetric reduced 
metric is homogeneous and we can solve completely the quantum Einstein 
equations for the minisuperspace model. 
In particular we have two regions (Region 1 and Region 2), the first one is where the matter is
localized (Region 1) with a Friedmann- Robertson-Walker type metric 
and the other one outside the matter (Region 2)
where the metric is of Kantowski-Sachs type \cite{KS}. 
We summarize the model in ADM variables following the paper of Nambu 
and Sasaki \cite{NS}.  After this we will pass to Ashtekar variables. 
In Ashtekar variables we quantize the system following ideas 
suggested from full loop quantum gravity \cite{book}.
The technology used in this paper was developed in the preview papers 
\cite{work1}, \cite{work2}, \cite{work3}, \cite{work4} and \cite{Bojk=1}. 

The paper is organized as follow.  In the first section we briefly
recall properties of the gravitational collapse in ADM variables inside
the event horizon, $r<2M G_N$, \cite{NS}.
As well known, here the temporal and
spatial (radial) coordinates exchange their role and 
so it is possible to study an homogeneous space-time.
In the second section we recall the Ashtekar formulation 
of general relativity in terms of the gauge connection $A_a^i$ and 
of the density triad $E^a_i$. In particular we recall the form 
of the Ashtekar connection and of the density triad inside the event 
horizon. For the symmetric reduced connection we introduce the 
holonomy and we define the classical Hamiltonian constraint in terms 
of holonomies and of the volume operator. At  the end of this section we 
define the boundary conditions on the $S^2$ separation sphere
(between Region 1 and Region 2 of above) in terms of holonomies.  
In the third section we recall the polymer quantization scheme 
and we introduce the kinematical Hilbert space for the gravitational
collapse. In the forth section we study the dynamics and we 
solve the Hamiltonian constraint inside and outside the matter.

\section{Gravitational collapse in ADM variables inside a black hole (classical theory)}
In this section we recall the results of \cite{NS} in ADM variables.
We consider the space-time region inside the black hole horizon 
and we study the collapse of a dust sphere. The metric inside the
matter (Region 1)($0 \leqslant \chi \leqslant \chi_0$) is described by 
a closed Friedmann universe homogeneous 
and isotropic,
\begin{eqnarray}
ds_{1}^2 = - N_1^2(t) dt^2 + R^2(t)(d \chi^2 + \sin^2 \chi (d \theta^2 + \sin^2 \theta d \phi^2)).
\label{IN}
\end{eqnarray}
The metric outside the collapsing star (Region 2)($\chi_0 \leqslant \chi < \infty$)
is of Kantowski-Sachs type 
\begin{eqnarray}
ds_{2}^2 = - N_2^2(t)dt^2 + a^2(t) d \chi^2 + b^2(t) (d \theta^2 + \sin^2 \theta d \phi^2).
\label{OUT}
\end{eqnarray}
We have assumed the 3-surface is described by the same radial
coordinate $\chi$ also in the exterior of the dust matter region.  


\begin{figure}
 \begin{center}
  \includegraphics[height=7cm]{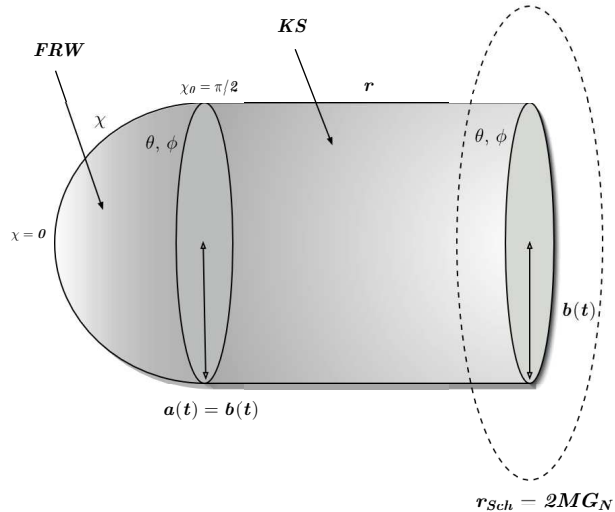}
  \end{center}
  \caption{\label{FRWKS} Spatial section for the gravitational collapse inside the event horizon
  for the particular value of the radial coordinate $\chi_0 = \pi/2$.}
  \end{figure}

The action in ADM variables using (\ref{IN}) and (\ref{OUT}) is
\begin{eqnarray}
&& S = \frac{1}{2 \kappa} \int d^3 x N \sqrt{h} (K^a_b K^a_b - (K^a_a)^2 + ^{(3)}\hspace{-0.1cm}{\rm R} - 2 \kappa \rho_{\rm dust}) = \nonumber \\
&& \hspace{0.4cm} = \frac{{\rm V}_1 \,  R^3}{2 \kappa} 
\Big[- \frac{6}{N_1} \Big(\frac{\dot{R}}{R}\Big)^2 + N_1\Big(\frac{6}{R^2} - 2 \kappa \rho_{{\rm dust}}\Big)\Big] + \nonumber \\
&& \hspace{0.4cm} + \frac{V_2 \, a b^2}{2 \kappa} \Big[ - \frac{1}{N_2} 
\Big(2 \, \Big(\frac{\dot{b}}{b} \Big)^2 + 4 \, \frac{\dot{a} \dot{b}}{a b} \Big) + \frac{2 N_2}{b^2} \Big],
\end{eqnarray}
where ${\rm V}_1 = 4 \pi \int_0^{\chi_0} d \chi \sin^2 \chi$,  
${\rm V}_2 \equiv 4 \pi \mathcal{L} \int_{\chi_0}^{\chi_1} d \chi \equiv 4 \pi \mathcal{L} (\chi_1- \chi_0)$
 and $\kappa = 8 \pi G_N$ (the index $a,b = 1, 2, 3$). 
I introduced $\mathcal{L}$ for dimensional reason ($[\mathcal{L}] = L$) in $\rm{V}_1$, and $\chi_1$ is a cut-off on the space radial coordinate. The spatial homogeneity enable us to fix a linear radial cell and restrict all 
integrations to this cell \cite{Boj}.
To simplify notations we restrict the linear radial cell to $\chi_1 - \chi_0 = 1$.

The corresponding Hamiltonian is 
\begin{eqnarray}
&& H = p_R \dot{R} + p_a \dot{a} + p_b \dot{b} - L = \nonumber \\
&& \hspace{-2cm} = N_1 \Big( - \frac{\kappa}{12 {\rm V}_1} \frac{p_R^2}{R} - \frac{3 {\rm V}_1}{\kappa} R + M_{{\rm dust}} \Big) + 
N_2 \Big[\frac{2 \kappa}{{\rm V}_2} \Big( - \frac{p_a p_b}{4 b} + \frac{a \, p_a^2}{8 b^2}\Big)
- \frac{{\rm V}_2}{\kappa} a \Big] ,
\end{eqnarray}
where $M_{{\rm dust}} = {\rm V}_1 R^3 \rho_{{\rm dust}}$ is the constant total dust matter
and the momentum conjugate to the 3-metric $q_{ab}$ is 
\begin{eqnarray}
\Pi^a_b = - \frac{1}{2 \kappa} \, (K^a_b - K h^a_b) = \left\{ \begin{array}{ll} 
- \frac{1}{2 \kappa N_1} \sin \theta \, 
{\rm diag} \Big(2 \dot{R} \sin^2 \chi, \, 2 \dot{R}, \, 
2 \frac{\dot{R}}{\sin^2 \theta}  \Big) & {\rm if} \,\, \chi < \chi_0 ; \\
                                   - \frac{1}{2 \kappa N_2} \sin \theta \, {\rm diag}\Big(\frac{2 b \dot{b}}{a}, 
                                   \, \dot{a} + \frac{a \dot{b}}{b}, \, \frac{a \dot{b}}{b \sin^2 \theta} \Big) 
                                   &  {\rm if} \,\, \chi > \chi_0. \end{array} \right.
\end{eqnarray} 
The momentum conjugate to the variables $R, a, b$ are 
\begin{eqnarray}
p_R = - \frac{6 {\rm V}_1}{\kappa N_1} R \dot{R} \, , \hspace{0.6cm} 
p_a = - \frac{2 {\rm V}_2}{\kappa N_2} b \dot{b} \, , \hspace{0.6cm}
p_b = - \frac{2 {\rm V}_2}{\kappa N_2} ( a \dot{b}+ b \dot{a}) \, .
\end{eqnarray}
To obtain a correlation between Region 1 and Region 2 we have to take 
into account the momentum constraint on the $S^2$-sphere junction surface 
that is in $\chi = \chi_0$ localized. 
When we impose the condition $\Pi^{\chi j}\,_{| j} =0$ on the surface in $\chi = \chi_0$ 
we obtain the following unambiguous 
junction conditions
\begin{eqnarray}
&& P_1 = \frac{p_R \sin^2 \chi_0}{3 {\rm V}_1} - \frac{p_a}{{\rm V}_2} = 0,  \nonumber \\
&& P_2 = R \sin \chi_0 - b = 0.
\label{CONSTR}
\end{eqnarray}
We can express the first of the relations (\ref{CONSTR}) in terms of $R, \dot{R}, b, \dot{b}$ and
we obtain the new constraints set 
\begin{eqnarray}
&& P_1 = \frac{R \dot{R} \sin^2 \chi_0}{N_1} - \frac{b \dot{b} }{N_2} = 0,  \nonumber \\
&& P_2 = R \sin \chi_0 - b = 0.
\label{CONSTR2}
\end{eqnarray}

The gravitational collapse inside the horizon in ADM variables is completely defined by the following four constraints \cite{NS}
\begin{eqnarray}
&& H_1 = - \frac{\kappa}{12 {\rm V}_1} \frac{p_R^2}{R} - \frac{3 {\rm V}_1}{\kappa} R + M_{{\rm dust}} = 0 , \nonumber \\
&& H_2 = \frac{2 \kappa}{{\rm V}_2} \Big( - \frac{p_a p_b}{4 b} + \frac{a \, p_a^2}{8 b^2}\Big)
- \frac{{\rm V}_2}{\kappa} a = 0,\nonumber  \\
&& P_1 = \frac{p_R \sin^2 \chi_0}{3 {\rm V}_1} - \frac{p_a}{{\rm V}_2} = 0,  \nonumber \\
&& P_2 = R \sin \chi_0 - b = 0.
\label{CONSTRAINTADM}
\end{eqnarray}
Solving the constraints equations (\ref{CONSTRAINTADM}) we obtain the known results 
for the classical dust matter gravitational collapse \cite{NS}.

  
  \begin{figure}
 \begin{center}
  \includegraphics[height=5cm]{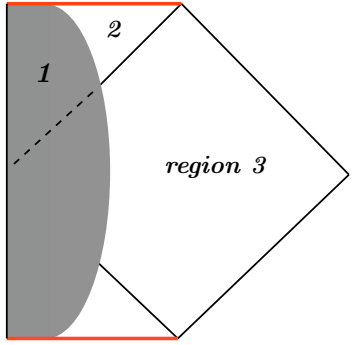}
  \end{center}
  \caption{\label{PCol} Penrose diagram for the gravitational collapse inside the event horizon (Region 1 and region 2) and outside the event horizon (Region3).}
  \end{figure}

\section{Gravitational collapse in Ashtekar variables}
In this section we study the gravitational collapse in Ashtekar variables \cite{variables}.
In particular we will express the Hamiltonian constraint inside and outside the
matter and the constraints $P_1$ and $P_2$ in terms of the symmetric reduced 
Ashtekar connection \cite{Bojk=1}, \cite{BojThiemann}.

\subsection{Ashtekar variables}

In LQG the fundamental variables are the Ashtekar variables: they consist  
of an $SU(2)$ connection $A_a^i$ and the electric field $E_i^a$, where $a, b, c, \dots = 1, 2, 3$ are tensorial indices on the spatial section and $i, j, k, \dots = 1, 2, 3$ are indices in the 
$su(2)$ algebra.   
The density weighted triad $E^a_i$ is related to the triad $e^i_a$ by the 
relation $E^a_i = \frac{1}{2} \epsilon^{abc} \, \epsilon_{ijk} \, e^j_b \, e^k_c$.  The 
metric is related to the triad by $q_{a b} = e^i_a \, e^j_b \, \delta_{ij}$.
Equivalently, 
\begin{equation}
\sqrt{\mbox{det}(q)} \, q^{a b} = E^a_i \, E^b_j \, \delta^{ij}.    
\end{equation}
The rest of the relation between the variables $(A^i_a, E^a_i)$ and the ADM variables 
$(q_{ab}, K_{ab})$ is given by
\begin{eqnarray}
A^i_a = \Gamma^i_a + \gamma \, K_{ab} E^b_j \delta^{i j}
\end{eqnarray} 
where $\gamma$ is the Immirzi parameter and $\Gamma^i_a$ is the spin 
connection of the triad, namely the solution of Cartan's equation: 
$\partial_{[a} e^i_{b]} + \epsilon^i_{j k} \, \Gamma^j_{[a} e^k_{b]} = 0$.

The action is 
\begin{eqnarray}
S = \frac{1}{\kappa \,\gamma} \int dt \int_{\Sigma} d^3 x 
\left[- 2 \mbox{Tr}(E^a \dot{A}_{a}) - N \mathcal{H} - N^a \mathcal{H}_a - N^i \mathcal{G}_i \right],
\label{action}
\end{eqnarray}  
where $N^a$ is the shift vector, $N$ is the lapse function and $N^i$ is the Lagrange
multiplier for the Gauss constraint $\mathcal{G}_i$. 
We have introduced also the notation $E_{[1]} = E^a \partial_a = E^a_i \tau^i \partial_a$
and $A_{[1]} = A_{ a} d x^a = A^i_a \tau^i d x^a$.
The functions $\mathcal{H}$, 
$\mathcal{H}_a$ and $\mathcal{G}_i$
are respectively the Hamiltonian, diffeomorphism and Gauss constraints, given by
\begin{eqnarray}
&& \mathcal{H}(E^a_i, A^i_a) = 
- 4 \, e^{-1} \, \mbox{Tr} \left(F_{ab} \, E^a E^b \right) 
- 2 \, e^{-1} \, (1+ \gamma^2) \, E^a_i E^b_j K^i_{[a} K^j_{b]}
\nonumber \\
&& \mathcal{H}_b(E^a_i, A^i_a) = E^a_j \, F^j_{ab} - (1+ \gamma^2) K^i_a G_i \nonumber \\
&& \mathcal{G}_i(E^a_i, A^i_a) = \partial_a E^a_i + \epsilon_{ij}^k \, A^j_a E^a_k, 
\label{constraints}
\end{eqnarray}
where the curvature field strength is 
$F_{ab} = \partial_a A_{ b} - \partial_b A_{a} + \left[A_{ a}, A_{b} \right]$ and the 
determinant is 
$e = \mbox{det}(e^i_a) \equiv \sqrt{|\mbox{det}(E_{[1]})} \, \mbox{sgn}(\mbox{det}(E_{[1]})$ 
\cite{MAT}.

\subsection{The Hamiltonian constraint inside the matter}

In this section we recall the Ashtekar variables for an homogeneous and
isotropic space-time of topology $R \times S^3$ \cite{Bojk=1}. For this space the Ashtekar's 
connection and the densitized triad are 
\begin{eqnarray}
A^i_a = c \, \omega^I_a\, \delta^i_I \, , \,\,\,\,\,\,\, E^a_i = p X^a_I \, \delta_i^I,
\end{eqnarray}
where $\omega^I$ are the left-invariant one-forms and $X_I$ are the 
left-invariant vector fields fulfilling $\omega^I(X_J) = \delta^I_J$.
The Hamiltonian constraint in terms of the variables $(c, p)$ is 
\begin{eqnarray}
H = H^{(RW)}_{E} + H_M = \frac{12}{\kappa} \, c \, (1 - c) \mbox{sgn}(p) \sqrt{|p|} + H_M
\end{eqnarray}
where $H^{(RW)}_{E}$ is the Euclidean part of the Hamiltonian constraint inside the matter and 
$H_M = M_{\rm dust}$ is the Hamiltonian constraint for the dust matter.  
I recall that the metric of the space-time in the region where the matter is localized is 
\begin{eqnarray}
ds^2= -dt^2 + a^2(t)(d \chi^2 + \sin^2 \chi(\sin^2 \theta d \phi^2 + d \theta^2)).
\end{eqnarray}
The volume operator for the space section inside the matter in Ashtekar variables is
\begin{eqnarray}
V^{(RW)} =  2 \pi (\chi_0 - \cos \chi_0 \sin \chi_0)  |p|^{\frac{3}{2}} \equiv {\rm V}(\chi_0) |p|^{\frac{3}{2}}. 
\end{eqnarray} 
The classical theory is define by the symplectic structure $\{c, p\} = \gamma \kappa /3$ \cite{Bojk=1}.

\paragraph{Holonomies and the Hamiltonian constraint inside the matter.}

We introduce the holonomies for a space-time homogeneous and isotropic
in the direction $I$
\begin{eqnarray}
h_I^{(RW)} = e^{\delta_0 c \tau_I} = \cos(c \, \delta_o/2) + 2 \, \tau_I \sin(c \, \delta_0/2),
\label{holIn}
\end{eqnarray}
and we express the gravitational part of the Hamiltonian constraint in terms of the holonomies 
\begin{eqnarray}
H^{(RW)}_{E} =  - \frac{8}{\kappa^2 \gamma \delta_0^3 \rm{V}(\chi_0)} 
\sum_{IJK} \mbox{Tr}\big[h_I^{(RW)} h_J^{(RW)} h_I^{(RW)-1} h_J^{(RW)-1} h^{(RW)-1}_{[I,J]} h_K^{(RW)} \{h_K^{(RW) -1}, V^{(RW)} \}\big].
\label{HamInside}
\end{eqnarray}

\subsection{The Hamiltonian constraint outside the matter}
We recall that outside the matter but inside the horizon we have
a Kantowski-Sachs type space-time. 
An homogeneous but anisotropic space-time of spatial section $\Sigma$
of topology $\Sigma \cong \mathbb{R} \times S^2$ is characterized by an invariant 
connection 1-form $A_{[1]}$ of the form \cite{BojThiemann}, \cite{BojImp}
\begin{eqnarray}
A_{[1]} = A(t) \, \tau_3 \, dr + (A_1(t) \, \tau_1 + A_2(t) \tau_2) \, d \theta +
           (A_1(t) \, \tau_2 - A_2(t) \tau_1) \sin \theta \, d \phi + \tau_3 \, \cos \theta \, d \phi.
           \label{symconnection}
\end{eqnarray} 
The $\tau_i$ are the generators of the $SU(2)$ fundamental representation. 
They are related to the
Pauli $\sigma_i$ matrix by $\tau_i = -  \frac{i}{2} \sigma_i$.
On the other side the dual invariant densitized triad is 
\begin{eqnarray}
E_{[1]} = E(t) \, \tau_3 \, \sin \theta \, \frac{\partial}{\partial r} + (E^1(t) \, \tau_1 + E^2(t) \, \tau_2)
\, \sin \theta \, \frac{\partial}{\partial \theta}  +
           (E^1(t) \, \tau_2 - E^2(t) \tau_1) \frac{\partial}{\partial \phi}.
           \label{symtriad}
\end{eqnarray} 
The coordinate $r$ is related to the coordinate $\chi$ used in the first section by 
$r = l_P \chi$.

We are interested to the Kantowski-Sachs space-time 
with space section of topology $\mathbb{R} \times S^2$; the connection $A_{[1]}$ 
is more simple than in (\ref{symconnection}) with $A_2 = A_1$, and in the triad 
(\ref{symtriad}) we can choose the gauge $E^2 = E^1$ \cite{Bombelli}. There is 
a residual gauge freedom on the pair $(A_1, E^1)$. This is a discrete transformation
$P: (A_1, E^1) \rightarrow (-A_1, -E^1)$; we have to fix this symmetry on the Hilbert space.
The Gauss constraint is automatically satisfied and the Euclidean part of the 
Hamiltonian constraint is 
\begin{eqnarray}
H_E = \frac{8 \pi \,  l_P \sqrt{2} \,\mbox{sgn}(E)}{\sqrt{|E|}|E^1|} \, 
\Big[2 A E A_1 E^1 + ( 2 (A_1)^2  - 1)(E^1)^2 \Big].
 \label{hamiltonianE}
\end{eqnarray}

We recall that the relation between the metric and density triad formulation 
and that the volume of the space section $\Sigma$ are 
\begin{equation}
q_{a b} = \mbox{diag}\Big(\frac{2 (E^1)^2}{|E|}, |E|,  |E| \, \sin^2 \theta\Big) \,\, , \,\,\,\,\,
V  = 4 \pi \sqrt{2} l_P \sqrt{|E|} |E^1|.
\end{equation}

The phase space consists of two canonical  pairs $A, E$ and $A_1, E^1$ and 
the simplectic structure is given by 
the poisson brackets, $\{A, E\} = \frac{\kappa \gamma}{4 \pi l_P}$ and 
$\{A_1, E^1\} = \frac{\kappa \gamma}{16 \pi  l_P}$  \cite{work4} 
The coordinates and the momenta have dimensions: $[A] = L^{-1}$, $[A_1] = L^0$,
$[E] = L^2$ and $[E^1] = L$. 

\paragraph{Holonomies and Hamiltonian constraint outside the matter.}

We introduce the fiducial triad $^o e^a_I = \mbox{diag}(1, 1, \sin^{-1} \theta)$ and co-triad 
$^o \omega^I_a = \mbox{diag}(1, 1, \sin \theta)$, and define the holonomy : 
$h = \mbox{e}^{\int A_{[1]}} = \mbox{e}^{ \int A_{[1] a} d x^a} = \mbox{e}^{\int A^i_a \, \frac{d x^a}{d \lambda} \, \tau_i d \lambda}
= \mbox{e}^{\int A^i_I  \,^o\omega^I_a \,^oe_J^a  \, u^J \, \tau_i d \lambda} = 
\mbox{e}^{\int A^i_I  \, u^I \, \tau_i d \lambda}$,
where $u^a = \frac{d x^a}{d \lambda} = (\frac{dr}{d \lambda}, \frac{d \theta}{d \lambda}, \frac{d \phi}{d \lambda})$ and $u^I =  \,^o\omega^I_a \, u^a$.
The holonomy along a curve
in the direction ``$I \, $" 
is given by
\begin{eqnarray}
&& h_1 = \exp \int A^i_1 \tau_i d x^1 = \exp[ A \mu_0 l_P \, \tau_3] , \nonumber \\
&& h_2 = \exp \int A^i_2 \tau_i d x^2 = \exp [A_1 \mu_0  \, (\tau_2 + \tau_1)] , \nonumber \\
&& h_3 = \exp \int A^i_3 \tau_i d x^3 = \exp [A_1 \mu_0   \, (\tau_2 - \tau_1) ], 
\label{holonomiyI}
\end{eqnarray}
where $A_1^i = (0, 0, A)$, $A_2^i = (A_1, A_1,0)$ and $A_3^i = (-A_1, A_1, 0)$.
The connection in (\ref{holonomiyI}) is integrated in the direction ``$I \, $";
$\mu_0 l_P$ is the length of the curve along the direction $r$, 
$\mu_0$ is the length of the curve along the directions $\theta$ and  $\phi$ \cite{MAT}
\footnote{In the reference \cite{MAT} the authors obtain $\mu_0 \sim 1$.
}.
The length are defined using the fiducial triad  $^o e^a_I$.
Introducing the normalized vectors $n_1^i = (0, 0, 1)$, $n_2^i = \frac{1}{\sqrt{2}} (1, 1, 0)$,
  $n_3^i = \frac{1}{\sqrt{2}} (-1, 1, 0)$ we can rewrite the holonomy $h_I$ in the
  direction  ``$I \, $" as
  \begin{eqnarray}
&& h_I= \exp \left( \bar{A}_I \, n^i_I  \, \tau_i \right)= \cos\Big(\frac{\bar{A}_I}{2}\Big) + 2 n^i_I \, \tau_i \, \sin\Big(\frac{\bar{A_I}}{2}\Big)  , \nonumber \\
&& h_1 = \cos\Big(\frac{A \mu_0 l_P}{2}\Big) +  2 \tau_3 \, \sin\Big(\frac{A \mu_0 l_P}{2}\Big)  , \nonumber \\
&& h_2 = \cos\Big(\frac{\sqrt{2} A_1 \mu_0}{2}\Big) +\sqrt{2} (\tau_2 + \tau_1)\, \sin\Big(\frac{\sqrt{2} A_1 \mu_0}{2}\Big)  , \nonumber \\
&& h_3 =  \cos\Big(\frac{\sqrt{2} A_1 \mu_0}{2}\Big) +\sqrt{2} (\tau_2 - \tau_1) \, \sin\Big(\frac{\sqrt{2} A_1 \mu_0}{2}\Big),
\label{holonomiyI2}
\end{eqnarray}
where $\bar{A}_{I=1} = A l_P \mu_0$ and $\bar{A}_{I=2} = \bar{A}_{I=3} = A_1 \mu_0 \sqrt{2}$.


We now write the hamiltonian constraint (\ref{hamiltonianE}) in terms of holonomies
\begin{eqnarray}
 H_E  =  - \frac{16 \pi}{\kappa \gamma \mu_0^3} \, \sum_{I J K}\, \epsilon^{I J K} \, \mbox{Tr} \left[h_I h_J h_I^{-1} h_J^{-1} h_{[IJ]} \, h_K^{-1} \{h_K, V \}  \right],
\label{hamiltonianEreg2}
\end{eqnarray}
where 
 $h_{[IJ]} = \exp(- \mu_0^2 \, C_{I J} \, \tau_3) = \cos(\mu_0^2 \, C_{IJ}/2) - 2 \, \tau_3 \, \sin
 (\mu_0^2 \, C_{IJ}/2)$  
and $C_{IJ} = \delta_{2 I} \delta_{3 J} - \delta_{3_I} \delta_{2 J}$.


\subsection{Boundary conditions in Ashtekar variables}\label{Boun} 
In this section we recall relations between the Ashtekar and the ADM 
variables. In Region 1 where the matter is localized the relations are \cite{BojRP} 
\begin{eqnarray}
c = \frac{1}{2}\Big(1 - \gamma \frac{\dot{a}}{N_1}\Big) \, , \hspace{0.5cm} |p| = a^2
\label{Rel1}
\end{eqnarray}
In Region 2 instead the relations can be obtained comparing the Hamiltonian constraint
written in terms of ADM variables and in terms of Ashtekar
variables. From this analysis we obtain 
\begin{eqnarray}
A =  i \frac{\dot{a}}{N_2}  \, , \hspace{0.5cm} A_1 = \frac{i \, \dot{b}}{\sqrt{2} \, N_2} \, \hspace{0.5cm} |E| = b^2 \, \hspace{0.5cm} (E^1)^2 = \frac{a^2 b^2}{2}.
\label{Rel2}
\end{eqnarray}
Introducing (\ref{Rel1}) and (\ref{Rel2}) in (\ref{CONSTR2}) obtaining  
\begin{eqnarray}
\sqrt{|p|} \, (2 c -1) \sin^2 \chi_0 - \sqrt{2 |E|} \, A_1 = 0.
\label{ConstrAV}
\end{eqnarray}
We can express (\ref{ConstrAV}) in terms of holonomies using 
(\ref{holIn}) and (\ref{holonomiyI})  and we obtain  
\begin{eqnarray}
\frac{1}{\delta_0} \, \sqrt{|p|} \, [4 {\rm Tr} (h_1^{(RW)} \tau_1) + \delta_0] \, \sin^2 (\chi_0)- \frac{1}{\mu_0} \, \sqrt{2 |E|} \, 
{\rm Tr} [h_2 (\tau_2 + \tau_1)] + O(\mu_0) = 0.
\label{ConstrAVH}
\end{eqnarray}
Introducing the explicit form of holonomies in (\ref{ConstrAVH}) we 
obtain the following finally form for the boundary conditions (\ref{CONSTR2})
  \begin{eqnarray}
&& \frac{1}{\delta_0} \, \sqrt{|p|} \, 
\Big[4 \sin\Big(\frac{c \, \delta_0}{2}\Big) + \delta_0 \Big] \, \sin^2 (\chi_0)
- \frac{1}{\mu_0} \, 2 \sqrt{2 |E|} \, 
\sin\Big(\frac{A_1 \, \mu_0}{\sqrt{2}}\Big) + O(\mu_0) = 0, \nonumber \\
&& \sqrt{|p|} \, \sin(\chi_0) - \sqrt{|E|} = 0.
\label{ConstrAVH2}
\end{eqnarray}

\section{Quantum kinematics}
We want to quantize the collapse inside the horizon  
using techniques from loop quantum gravity. 
Now we build the kinematical Hilbert space $\mathcal{H}_{kin}$. 
We define a graph $\Gamma$ as a countable number of triple of points 
$(c_i, \mu_{E i} \, , \, \mu_{E^{1} i})$, where $c_i, \mu_{E i}, \mu_{E^1i} \in \mathbb{R}$.
We denote by $\mbox{Cyl}_{\Gamma}$ the vector space of functions $f(c, A, A_1)$ 
($f : \mathbb{R}^3 \rightarrow  \mathbb{C}$) of the type 
\begin{eqnarray}
f(c, A, A_1) = \sum_{i \, j, k} f_{ijk} \,
e^{\frac{i \mu_i \, c}{2} + \frac{i \mu_{E j} \, l_P \, A}{2} + \frac{i \mu_{E^1 k} \, A_1}{\sqrt{2}}}.
\label{genfun}
\end{eqnarray}
where $c, A, A_1 \in \mathbb{R}$, $c_i, \mu_{E j} \, , \, \mu_{E^1 k} \in \mathbb{R}$ ,
$f_{ijk} \in \mathbb{C}$ and 
$i,j,k$ run over a finite number of integers (labelling the points of the graph). We call
the function $f(c, A, A_1)$ in Cyl$_{\Gamma}$ cylindrical  with respect to the graph $\Gamma$.
We consider all possible graphs (the points and their number can vary 
from a graph to another) and denote by Cyl the infinite dimensional 
vector space of functions cylindrical with respect to some graph: $\mbox{Cyl}=\bigcup_{\Gamma}\mbox{Cyl}_{\Gamma}$. 
Thus, any element $f(c, A, A_1)$ of Cyl can be expanded as in (\ref{genfun}),
where the uncountable basis 
$e^{\frac{i \mu_{i} \, c}{2}} \otimes 
e^{\frac{i \mu_{E i} \, l_P \, A}{2}} \otimes e^{\frac{i \mu_{E^1 j} \, A_1}{\sqrt{2}}}$
is now labeled by arbitrary real numbers $(c, \mu_{E} \, , \, \mu_{E^{1}})$.
A basis in Cyl 
is given by $|\mu, \mu_E,  \mu_{E^1} \rangle \equiv |\mu \rangle \otimes |\mu_E \rangle \otimes |\mu_{E^1} \rangle$.  
Introducing the standard bra-ket notation we can define a basis \cite{MAT} in the
Hilbert space via  
\begin{eqnarray}
\langle c |\mu \rangle \otimes \langle A|\mu_E\rangle \otimes \langle A_1|\mu_{E^1} \rangle =  
e^{\frac{i \mu \, c}{2}} \otimes
e^{\frac{i \mu_E \, l_P \, A}{2}} \otimes e^{\frac{i \mu_{E^1} \, A_1}{\sqrt{2}}}.
\label{basis}
\end{eqnarray} 
The basis states (\ref{basis}) are normalizable in contrast to the standard quantum mechanical 
representation and they satisfy 
\begin{eqnarray}
\langle \mu, \mu_E, \mu_{E^1}| \nu, \nu_E, \nu_{E^1} \rangle = \delta_{\mu, \nu} \,\delta_{\mu_E, \nu_E} \, \delta_{\mu_{E^1}, \nu_{E^1}}.
\end{eqnarray}
The Hilbert space $\mathcal{H}_{kin}$ is the Cauchy completion of Cyl or more succinctly $\mathcal{H}_{kin} = L_2(\bar{\mathbb{R}}^3_{Bohr}, d \mu_0)$, where $\bar{\mathbb{R}}_{Bohr}$ is the Bohr-compactification of $\mathbb{R}$ and $d\mu_0$ is the Haar measure on $\bar{\mathbb{R}}^3_{Bohr}$.

We can quantize the theory using the standard quantization procedure 
$\{\, , \,\} \rightarrow - \frac{i}{\hbar} [\, , \, ]$. 
We recall the fundamental difference from the standard Schr$\ddot{\mbox{o}}$dinger 
quantization program.  
In the ordinary Schr$\ddot{\mbox{o}}$dinger 
representation of the Weyl-Heisenberg algebra the classical fields $c$, $A$ and $A_1$ 
translate in operators. In loop quantum gravity on the contrary the operators $c$, $A$ and $A_1$ do not exist \cite{LoopOld}.
We can not promote the Poisson brackets to commutators $[\hat{c}, \hat{p}] = i \gamma l_P^2/3$, 
$[\hat{A}, \hat{E}] = i \gamma l_P/4 \pi$ and $[\hat{A}_1, \hat{E}^1] = i \gamma l_P/16 \pi$; 
rather, the quantum fundamental operators are $\hat{c}, \hat{E}, \hat{E}^1, \hat{h}_I$.
The momentum operators can be represented on the Hilbert space by 
\begin{eqnarray}
\hat{p} \rightarrow - i \frac{\gamma \, l_P^2}{3} \, \frac{d}{d c} \, ,  \,\,\,\,\,\,\,\,
\hat{E} \rightarrow - i \frac{\gamma \, l_P}{4 \pi} \, \frac{d}{d A} \, ,  \,\,\,\,\,\,\,\,
\hat{E}^1 \rightarrow - i \frac{\gamma \, l_P}{16 \pi} \, \frac{d}{d A_1}.
\label{EE1}
\end{eqnarray}
It is easy to calculate the spectrum of these two momentum operators on the Hilbert space 
basis. This is given by 
\begin{eqnarray}
&& \hat{p} |\mu, \mu_E, \mu_{E^1} \rangle = \frac{\mu \,  \gamma \, l_P^2}{6} |\mu, \mu_{E}, \mu_{E^1} \rangle , \nonumber \\
&& \hat{E} |\mu, \mu_E, \mu_{E^1} \rangle = \frac{\mu_E \,  \gamma \, l_P^2}{8 \pi} |\mu, \mu_{E}, \mu_{E^1} \rangle \, ,
\,\,\,\,\,\,\,\, 
\hat{E}^1 |\mu, \mu_E, \mu_{E^1} \rangle = \frac{\mu_{E^1} \, \gamma \, l_P}{16 \pi \sqrt{2}} |\mu, \mu_E, \mu_{E^1} \rangle.
\label{diagEE1}
\end{eqnarray}

 We have to fix the residual gauge freedom on the Hilbert space outside the matter. 
 We consider the
 operator $\hat{P} : |\mu, \mu_E, \mu_{E^1} \rangle \rightarrow |\mu, \mu_E, - \mu_{E^1} \rangle$
 and we impose that only the invariant states (under $\hat{P}$) are in the kinematical
 Hilbert space. The states in the Hilbert space are : 
 $\frac{1}{\sqrt{2}} \left[|\mu, \mu_E, \mu_{E^1} \rangle + |\mu, \mu_E, - \mu_{E^1} \rangle\right]$
 for $\mu_{E^1} \neq 0$ and the states $| \mu, \mu_E , 0 \rangle$ for $\mu_{E^1} = 0$.
 
 The holonomy operators $\hat{h}_{I}$ in the directions $r$, $\theta$, $\phi$ of the space 
 section $\mathbb{R} \times S^2$ outside the matter are : $\hat{h}_1^{(\mu_E)}$, $\hat{h}_2^{(\mu_{E^1})}$ and 
 $\hat{h}_3^{(\mu_{E^1})}$,
where $\mu_{E} l_P$ is the length along the radial direction $r$ 
and $\mu_{E_1}$ is the length along the directions $\theta$ and $\phi$
(all the length are define using the fiducial triad $^o e^a_I$). Inside the matter the 
holonomy operators are $\hat{h}_I^{(RW) (\mu)}$
where $\mu$ is the length along any direction $\chi$, $\theta$ or $\phi$ inside 
the matter. The holonomies 
operators act on the Hilbert space $\mathcal{H}_{kin}$ by multiplication.

\section{Hamiltonian constraint and quantum dynamics}
In this section we are going to study the dynamics of the model.
The general strategy to quantize a system with constraints was introduced by Dirac. 
To implement the Dirac strategy we must impose the classical constraints at
the quantum level to obtain the physical space. This strategy define the
quantum Einstein's equations that in general are define by (for $\gamma = i$)
\begin{eqnarray}
&& \hat{\mathcal{H}} | \psi \rangle =  
\left[- 4 \, e^{-1} \, \mbox{Tr} \left(F_{ab} \, E^a E^b \right) + \mathcal{H}_{M} \right] | \psi \rangle = 0,
\nonumber \\
&& \hat{\mathcal{H}}_b | \psi \rangle = \left[E^a_j \, F^j_{ab} + \mathcal{H}_{M b} \right]| \psi \rangle = 0, 
\nonumber \\
&& \hat{\mathcal{G}}_i |\psi \rangle = \left[\partial_a E^a_i + \epsilon_{ij}^k \, A^j_a E^a_k + \mathcal{G}_{M i} \right] | \psi \rangle = 0.
\label{Quanconstraints}
\end{eqnarray}

In the case of our minisuperspace (homogeneous) model the second and the third constraints are 
identically zero then to obtain the physical space we have to solve 
the first scalar constraint. In this paper we want to quantize the model
following the full loop quantum gravity ideas. To this aim we have expressed
the Hamiltonian constraint in terms of homogeneous holonomies and we have 
introduced the polymer representation of the Weyl algebra. 
At this point we have all the ingredients to solve the Hamiltonian
constraint inside and outside the region where the matter is localized and to
impose at quantum level the boundary condition introduced in section 
\ref{Boun}.

As in non-trivially constrained systems, we expect that the physical states are not 
normalizable in the kinematical Hilbert space. However, as in the full loop quantum 
gravity theory, we again have the triplet 
\begin{eqnarray}
\mbox{Cyl} \subset \mathcal{H}_{kin} \subset \mbox{Cyl}^*
\end{eqnarray}
of spaces and the physical states will be in $\mbox{Cyl}^*$, which is the algebraic dual of 
$\mbox{Cyl}$ \cite{book}.
A generic element of this space is 
\begin{eqnarray}
\langle \psi | = \sum_{\mu, \mu_E, \mu_{E^1}} \psi_{\mu, \mu_E, \mu_{E^1}} 
\langle \mu, \mu_E, \mu_{E^1} |.
\label{BHstate}
\end{eqnarray}

The quantum version of the Hamiltonian constraint outside 
the matter can be obtained promoting the classical holonomies
to operators and the poisson brackets to the commutators. 
Using the relations in (\ref{holonomiyI2}) we can express the quantum version of the 
Kantowski-Sachs Hamiltonian constraint (\ref{hamiltonianEreg2}) as
\begin{eqnarray}
&& \hat{H}_E  =  - \frac{16 \pi \, i}{\mu_0^3 \, \gamma \, l_P^2} \, 
 \Big[4 \sin(2x) \sin(2y) \left(\sin(y) \hat{V} \cos(y) - \cos(y) \hat{V} \sin(y) \right) +\nonumber \\
&&\hspace{-0.5cm} 
 +2 \left( \cos(\delta) \sin^2(2y) - \sin(\delta) \left(\sin^2(2y) + 2 \cos(2y)\right)\right)
\left(\sin(x) \hat{V} \cos(x) - \cos(x) \hat{V} \sin(x) \right) \Big],
 \label{hamiltonianEreg2Q2}
\end{eqnarray}
 where we have introduced the following notations: 
 $x = A \mu_0 l_P/2$, $y = \sqrt{2} A_1 \mu_0/2$ and $\delta = \mu_0^2/2$.

Using the exponential form for the trigonometric function 
 we can calculate the action of the Hamiltonian constraint 
on the Hilbert space basis (\ref{basis}) \cite{work4}. 
At this point we can solve the Hamiltonian constraint outside the matter 
to obtain a first relation for the coefficients $\psi_{\mu, \mu_E, \mu_{E^1}}$ in the (\ref{BHstate}).
The constraint equation $\hat{H}_E |\psi \rangle = 0$ is now interpreted as an equation in the dual space $\langle \psi | \hat{H}^{\dag}_E$;
from this equation we can derive a relation for the coefficients $\psi_{\mu, \mu_E, \mu_{E^1}}$  
\begin{eqnarray}
&& \hspace{0.25cm}(V_{\mu_E - 2 \mu_0, \mu_{E^1} -  3 \mu_0} - V_{\mu_E - 2 \mu_0, \mu_{E^1} -  \mu_0}) 
\, \psi_{\mu, \mu_E - 2 \mu_0, \mu_{E^1} - 2 \mu_0} \nonumber \\
&& + (V_{\mu_E + 2 \mu_0, \mu_{E^1} -  \mu_0} - V_{\mu_E + 2 \mu_0, \mu_{E^1} -  3 \mu_0} ) \, \psi_{\mu, \mu_E + 2 \mu_0, \mu_{E^1} - 2 \mu_0} \nonumber \\
&& + (V_{\mu_E - 2 \mu_0, \mu_{E^1} + 3 \mu_0} -  V_{\mu_E - 2 \mu_0, \mu_{E^1} +  \mu_0}) \, \psi_{\mu, \mu_E - 2 \mu_0, \mu_{E^1} + 2 \mu_0} \nonumber \\ 
&& + (V_{\mu_E + 2 \mu_0, \mu_{E^1} +  \mu_0} - V_{\mu_E + 2 \mu_0, \mu_{E^1} +  3 \mu_0}) \, \psi_{\mu, \mu_E + 2 \mu_0, \mu_{E^1} + 2 \mu_0} \nonumber \\
&& + \frac{\sin(\mu_0^2/2) -  \cos(\mu_0^2/2)}{2}
\Big[(V_{\mu_E + \mu_0, \mu_{E^1} - 4 \mu_0} - V_{\mu_E - \mu_0, \mu_{E^1} - 4 \mu_0} ) \, \psi_{\mu, \mu_E, \mu_{E^1} - 4 \mu_0} \nonumber \\
&& - 2 \, (V_{\mu_E + \mu_0, \mu_{E^1}} - V_{\mu_E - \mu_0, \mu_{E^1}}) \, \psi_{\mu, \mu_E, \mu_{E^1}} 
+ (V_{\mu_E + \mu_0, \mu_{E^1} + 4 \mu_0} - V_{\mu_E - \mu_0, \mu_{E^1} + 4 \mu_0}) \, \psi_{\mu, \mu_E, \mu_{E^1} + 4 \mu_0}\Big] \nonumber \\
&& - 2 \sin(\mu_0^2/2) \Big[(V_{\mu_E + \mu_0, \mu_{E^1} - 2 \mu_0} 
- V_{\mu_E - \mu_0, \mu_{E^1} - 2 \mu_0}) \, \psi_{\mu, \mu_E, \mu_{E^1} - 2 \mu_0} \nonumber \\
&& + (V_{\mu_E + \mu_0, \mu_{E^1} +2 \mu_0} - V_{\mu_E - \mu_0, \mu_{E^1} +2 \mu_0} ) \, \psi_{\mu, \mu_E, \mu_{E^1} +2 \mu_0}\Big]= 0,
\label{solution}
\end{eqnarray}
where the volume eigenvalues are defined by 
\begin{eqnarray}
&&\hat{V} |\mu, \mu_E, \mu_{E^1} \rangle = V_{\mu_E, \mu_{E^1}}|\mu, \mu_E, \mu_{E^1} \rangle, \nonumber  \\ 
&& V_{\mu_E, \mu_{E^1}} \equiv \frac{\gamma^{\frac{3}{2}} l_P^3}{4 \sqrt{8 \pi}} 
\sqrt{|\mu_E|} \, |\mu_{E^1}|.
\label{AutVol}
\end{eqnarray}

The other constraint that we must to impose on the state (\ref{BHstate}) is the Hamiltonian 
constraint inside the matter.  If we introduce the holonomies (\ref{holIn}) in (\ref{HamInside}) we obtain the following trigonometric form of the operator 
\begin{eqnarray}
&& \hat{H}^{(RW)}_{E} = \frac{6 (1 - i)}{{\rm V}(\chi_0) l_P^2 \gamma \mu_0^3 \kappa} \, 
e^{- i \frac{c \, \mu_0 (\mu_0 + 4)}{2}} 
\Big[ - i + 2 (1+ i) \, e^{i c \, \mu_0} + 2 i \, e^{2 i c \, \mu_0} 
+2 (1 + i) \, e^{3 i c \, \mu_0} - i \, e^{ 4 i c \, \mu_0}  \nonumber \\
&& \hspace{1.5cm}+ e^{i c \, \mu_0^2}
- 2 (1+ i) \, e^{i c \, \mu_0 (\mu_0 + 1)} - 2 \, e^{i c \, \mu_0( \mu_0+2)} 
- 2( 1+ i) \, e^{i c \, \mu_0(\mu_0 + 3)} + e^{i c \, \mu_0 (\mu_0+4)} \Big] \nonumber \\
&& \hspace{1.5cm} \Big[ \sin \Big(\frac{c \, \mu_0}{2} \Big) \, \hat{V}^{(RW)} \, \cos \Big(\frac{c \, \mu_0}{2} \Big) 
- \cos \Big(\frac{c \, \mu_0}{2} \Big) \, \hat{V}^{(RW)} \, \sin \Big(\frac{c \, \mu_0}{2} \Big) \Big],
\label{HamTrig}
\end{eqnarray}
(the operator $c$ is not defined but the exponential and trigonometric operators 
are defined on the Hilbert space). We have simplified the notation using the 
identification $\delta_0 = \mu_0$.

The action of the operator (\ref{HamTrig}) on the basis  (\ref{basis}) is 
\begin{eqnarray}
&& \hat{H}_E^{(RW)} |\mu, \mu_E, \mu_{E^1} \rangle =
 \frac{3 \, (1 + i)}{\gamma l_P^2 \kappa \mu_0^3} \, (V_{\mu + \mu_0} - V_{\mu - \mu_0}) \,
\big[- i \, |\mu - \mu_0(\mu_0 + 4), \mu_E, \mu_{E^1} \rangle \nonumber\\
&& + 2 \, (1+i) \, | \mu - \mu_0(\mu_0 + 2), \mu_E, \mu_{E^1} \rangle 
+ 2 \, i \, | \mu - \mu_0^2, \mu_E, \mu_{E^1} \rangle +
2 \, (1 + i) \, |\mu - \mu_0(\mu_0 - 2), \mu_E, \mu_{E^1} \rangle \nonumber \\
&& - i \, |\mu - \mu_0(\mu_0 - 4), \mu_E, \mu_{E^1} \rangle  
 + |\mu + \mu_0(\mu_0 - 4) , \mu_E, \mu_{E^1}\rangle 
- 2 \, (1 + i) \, \mu + \mu_0(\mu_0 - 2), \mu_E, \mu_{E^1} \rangle \nonumber \\
&& -2 \, | \mu + \mu_0^2 , \mu_E, \mu_{E^1} \rangle 
 - 2 \, (1 +i) \, |\mu + \mu_0(\mu_0 + 2), \mu_E, \mu_{E^1} \rangle
+ |\mu - \mu_0(\mu_0 + 4), \mu_E, \mu_{E^1} \rangle  \big].
\label{ActHamIn}
\end{eqnarray}

We have all the ingredients to calculate the action of the Hamiltonian 
constraint inside the matter on the state (\ref{BHstate}). From equation
$\langle \psi | (\hat{H}^{(RW)}_{E} + \hat{H}_M) = 0$ 
($\langle \psi | \hat{H}_M = \langle \psi | M_{\rm dust}, \, \forall \, \mu \neq 0$
and $\langle \psi | \hat{H}_M = 0$ for $m = 0$ \cite{BREW}) 
we obtain another recursive relation 
for the coefficients $\psi_{\mu, \mu_E, \mu_{E^1}}$ 
\begin{eqnarray}
&& - i \, \big(V_{\mu + \mu_0(\mu_0+5)} - V_{\mu + \mu_0(\mu_0 + 3)}\big) \, 
\psi_{\mu + \mu_0(\mu_0+4), \mu_E, \mu_{E^1}} \nonumber \\
&& + 2 \,  (1+i) \, \big(V_{\mu + \mu_0(\mu_0+3)} - V_{\mu + \mu_0(\mu_0 -1)}\big) \, 
\psi_{\mu + \mu_0(\mu_0+2), \mu_E, \mu_{E^1}} \nonumber \\
&& +2 \, i \, \big(V_{\mu + \mu_0(\mu_0+1)} - V_{\mu + \mu_0(\mu_0 -1)}\big) \, 
\psi_{\mu + \mu_0^2, \mu_E, \mu_{E^1}} \nonumber \\
&& + 2 \, (1 + i) \, \big(V_{\mu + \mu_0(\mu_0 - 2)} - V_{\mu + \mu_0(\mu_0 - 3)} \big) \, 
\psi_{\mu + \mu_0(\mu_0 - 2), \mu_E, \mu_{E^1}}  \nonumber \\
&& - i \, \big(V_{\mu + \mu_0(\mu_0 - 3)} - V_{\mu + \mu_0(\mu_0 - 4)} \big) \, 
\psi_{\mu + \mu_0(\mu_0 - 4), \mu_E, \mu_{E^1}} \nonumber \\
&& + \big(V_{\mu - \mu_0(\mu_0 - 5)} - V_{\mu - \mu_0(\mu_0 - 3)} \big) \, 
\psi_{\mu - \mu_0(\mu_0 - 4), \mu_E, \mu_{E^1}} \nonumber \\
&& - 2 \, (1+ i) \, \big(V_{\mu - \mu_0(\mu_0 - 3)} - V_{\mu - \mu_0(\mu_0 - 1)} \big) \, 
\psi_{\mu - \mu_0(\mu_0 - 2), \mu_E, \mu_{E^1}} \nonumber \\
&& - 2 \, \big(V_{\mu - \mu_0(\mu_0 - 1)} - V_{\mu - \mu_0(\mu_0 + 1)} \big) \, 
\psi_{\mu - \mu_0^2, \mu_E, \mu_{E^1}} \nonumber \\
&& -2 \, (1 + i) \, \big(V_{\mu - \mu_0(\mu_0 + 1)} - V_{\mu - \mu_0(\mu_0 + 3)} \big) \, 
\psi_{\mu - \mu_0(\mu_0 + 2), \mu_E, \mu_{E^1}} \nonumber \\
&& + \big(V_{\mu - \mu_0(\mu_0 + 3)} - V_{\mu - \mu_0(\mu_0 + 5)} \big) \, 
\psi_{\mu - \mu_0(\mu_0 + 4), \mu_E, \mu_{E^1}} \nonumber \\
&& + \frac{\gamma \, l_P^2 \kappa {\rm V}(\chi_0) \, \mu_0^3}{3(i+1)} \, M_{\rm dust} \, \psi_{\mu, \mu_E, \mu_{E^1}}= 0,
\label{EqDisIn}
\end{eqnarray}
where we have introduced the volume eigenvalue inside the matter 
\begin{eqnarray}
&& \hat{V}^{(RW)} |\mu, \mu_E, \mu_{E^1} \rangle = V_{\mu}  |\mu, \mu_E, \mu_{E^1} \rangle,
\nonumber \\
&& V_{\mu} \equiv {\rm V}(\chi_0) \, \Big(\frac{\gamma l_P^2}{6}\Big)^{\frac{3}{2}} |\mu|^{\frac{3}{2}}.
\end{eqnarray}

The other constraints to impose are the boundary conditions in (\ref{ConstrAVH2}). The first 
of the two conditions assumes the following operator form 
\begin{eqnarray}
\frac{\sqrt{|p|}}{\mu_0} \, \Big[4 \sin\Big(\frac{c \, \mu_0}{2}\Big) + \mu_0 \Big] \, \sin (\chi_0)
-  2 \, \frac{\sqrt{2 |E|}}{\mu_0} \, \sin\Big(\frac{A_1 \, \mu_0}{\sqrt{2}}\Big)  = 0,
\label{ConstrBound}
\end{eqnarray} 
and the action of the operator (\ref{ConstrBound}) on the state (\ref{BHstate}) impose 
another condition on the physical state for the gravitational collapse
\begin{eqnarray}
&& \left(\sqrt{\frac{|\mu - \mu_0|}{3}} \, \psi_{\mu - \mu_0, \mu_E, \mu_{E^1}} -
\sqrt{\frac{|\mu + \mu_0|}{3}} \, \psi_{\mu + \mu_0, \mu_E, \mu_{E^1}} - i \, 
\mu_0 \, \sqrt{\frac{|\mu|}{3}} \, \psi_{\mu, \mu_E, \mu_{E^1}} \right) \sin^2 (\chi_0)
\nonumber \\
&& \hspace {2.5cm}- \, \sqrt{\frac{|\mu_E|}{2 \pi}} \, \big(\psi_{\mu, \mu_E, \mu_{E^1} -\mu_0}  - \psi_{\mu, \mu_E, \mu_{E^1}+ \mu_0} \big) = 0.
\label{Constr1}
\end{eqnarray}
 
The last constraint to impose is the second of equations (\ref{ConstrAVH2}). When we apply 
this constraint on the candidate physical state (\ref{BHstate}) we obtain 
\begin{eqnarray}
&& \sum_{\mu, \mu_E, \mu_{E^1}} \, \Big[\sqrt{|p|} \, \sin(\chi_0) - \sqrt{|E|} \, \Big] \,
\psi_{\mu, \mu_E, \mu_{E^1}} \,  |\mu, \mu_E, \mu_{E^1} \rangle = \nonumber \\ 
&& = \sum_{\mu, \mu_E, \mu_{E^1}} \,
\Bigg[\sqrt{\frac{|\mu|}{3}} \, \sin(\chi_0) - \sqrt{\frac{|\mu_E|}{4 \pi}} \, \Bigg] \, 
\psi_{\mu, \mu_E, \mu_{E^1}}  \,  |\mu, \mu_E, \mu_{E^1} \rangle = 0.
\label{AreaMach}
\end{eqnarray}
From equation (\ref{AreaMach}) we obtain that the gravitational collapse 
wave function $\psi_{\mu, \mu_E, \mu_{E^1}}$ 
depends only on the two parameters $\mu$ and $\mu_{E^1}$. 
For $\sin^2(\chi_0)=3/4 \pi$ 
we obtain
\begin{eqnarray}
\psi_{\mu, \mu_E, \mu_{E^1}} \equiv \psi_{\mu, \mu, \mu_{E^1}}.
\label{LastCon}
\end{eqnarray}   
  
 We conclude this section summarizing the quantum dynamical results. 
 In  this section we have solved the quantum Einstein equations and we 
 have obtained that the gravitational collapse wave function must satisfy
 the difference recursive equations (\ref{solution}), (\ref{EqDisIn}), (\ref{Constr1}) and 
 then we must impose in the result the constraint (\ref{LastCon}). 
 An important result in our simplified analysis is that all the coefficients in the difference equations 
 (\ref{solution}), (\ref{EqDisIn}) and (\ref{Constr1}) are regular in $\mu = 0$ and 
 in $\mu_E =0$ where the classical Schwarzschild singularity is localized.
 We can conclude that we have a regular and natural evolution beyond the classical
 singular point. In the next paragraph we will show the consistency of the 
 difference equations obtained in this section.

 \subsection{
 Physical states
 }
 In the previous paragraph we have calculated four difference equations 
 (\ref{solution}), (\ref{EqDisIn}), (\ref{Constr1}) and (\ref{LastCon}) that 
 the wave function for the gravitational collapse must satisfy. Now we 
 analyze this system introducing the boundary conditions
 in the dynamics. We write the physical state $|\psi \rangle$ in a tensor product 
 form for the Region 1 and Region 2 
 \begin{eqnarray}
 \langle \psi | = \sum_{\mu} \varphi_{\mu} \langle \mu | \otimes   \sum_{\mu_E, \mu_{E^1}} \phi_{\mu_E, \mu_{E^1}} 
\langle \mu_E, \mu_{E^1} |.
\label{TensorState}
 \end{eqnarray}
At this point to simplify the problem we follow the following steps.
\begin{enumerate}
\item The first difference equation is (\ref{EqDisIn}), this equation gives a 
relation for the coefficients $\varphi_{\mu}$ introduced in (\ref{TensorState});
we recall (\ref{EqDisIn}) using the simplification $\mu_0 = 1$ and introducing integer 
values for the independent variable $\mu$, $\mu = m \, \mu_0 \equiv m$ ($m \in 2 \mathbb{Z}$
\footnote{It is possible repeat the same analysis for the components with $m \in 2 \mathbb{Z} + 1$})
\begin{eqnarray}
&& - i \, \big(V_{m + 6} - V_{m + 4}\big) \, 
\varphi_{m + 5} 
+ 2 \,  (1+i) \, \big(V_{m + 4} - V_{m}\big) \, 
\varphi_{m + 3} 
+2 \, i \, \big(V_{m + 2} - V_{m}\big) \, 
\varphi_{m + 1} \nonumber \\
&& + 2 \, (1 + i) \, \big(V_{m- 1} - V_{m - 2} \big) \, 
\varphi_{m - 1}  
 - i \, \big(V_{m - 2} - V_{m - 3} \big) \, 
\varphi_{m- 3} 
+ \big(V_{m + 4} - V_{m + 2} \big) \, 
\varphi_{m + 3} \nonumber \\
&& - 2 \, (1+ i) \, \big(V_{m + 2} - V_{m} \big) \, 
\varphi_{m + 1} 
- 2 \, \big(V_{m} - V_{m - 2} \big) \, 
\varphi_{m - 1} 
-2 \, (1 + i) \, \big(V_{m - 2} - V_{m - 4} \big) \, 
\varphi_{m - 3} \nonumber \\
&& + \big(V_{m - 4} - V_{m - 6} \big) \, 
\psi_{m - 5} + \frac{\gamma \, l_P^2 \kappa {\rm V}(\chi_0)}{3(i+1)} \, M_{\rm dust} \, \varphi_m= 0,
\label{EqDisInm}
\end{eqnarray}
this difference equation can be solved for $\varphi_{m_0+5}$
introducing the following initial conditions: 
$\varphi_{m_0-5}, \, \varphi_{m_0-3}, \, \varphi_{m_0-1}, \, 
\, \varphi_{m_0+1}$ and $\varphi_{m_0+3}$ ($m_0 \in 2 \mathbb{Z}$).
The component $\varphi_{m_0}$ that is the even component in (\ref{EqDisInm})
can be calculated using the constraint (\ref{Constr1}) with $\mu_E = 0$
as we will stress at the point 3. of this paragraph.


\item The second equation is (\ref{solution}) with $\mu_0 =1$, 
$\mu_{E^1} = 2 n$ and $\mu_E = 2 l$ 
with $n, l  \in \mathbb{Z}$

\begin{eqnarray}
&& (V_{2 l - 2, 2 n -  3} - V_{2 l - 2, 2 n-  1}) 
\, \phi_{2l - 2, 2n - 2} 
+ (V_{2l + 2, 2n -  1} - V_{2l + 2, 2n -  3}) \, \phi_{2l + 2, 2n - 2} \nonumber \\
&& + (V_{2l - 2, 2n + 3} -  V_{2l - 2, 2n + 1}) \, \phi_{2l - 2, 2n + 2} 
+ (V_{2l + 2, 2n+ 1} - V_{2l + 2, 2n +  3}) \, \phi_{2l + 2, 2n + 2} \nonumber \\
&& + (\sin(\mu_0^2/2) -  \cos(\mu_0^2/2))
\Big[(V_{2l + 1, 2n - 4} - V_{2l - 1, 2n - 4}) \, \phi_{2l, 2n - 4} \nonumber \\
&& - 2 \, (V_{2l + 1, 2n} - V_{2l - 1, 2n}) \, \phi_{2l, 2n} 
+ (V_{2l + 1, 2n + 4} - V_{2l - 1, 2n + 4}) \, \phi_{2l, 2n + 4}\Big] /2\nonumber \\
&& - 2 \sin(\mu_0^2/2) \Big[(V_{\mu_E + \mu_0, \mu_{E^1} - 2 \mu_0} 
- V_{2l - 1, 2n - 2}) \, \phi_{2l, 2n - 2} \nonumber \\
&& + (V_{2l + 1, 2n +2} - V_{2l - 1, 2n +2} ) \, \phi_{2l, 2n +2}\Big]= 0.
\label{solutionln}
\end{eqnarray}
We can solve this difference equation to obtain $\phi_{2l, 2n}$
giving the boundary conditions :
$\phi_{- 2 l_0, 2n}$, $\phi_{2 l_0, 2n}$ $\forall n$ and 
$\phi_{2l, 2n_0 - 4}$,  $\phi_{2l, 2n_0 -2}$, $\phi_{2l, 2n_0}$, $\phi_{2l, 2n_0+2}$
$\forall l$.

\item The last equation is (\ref{Constr1}) with $\sin^2(\chi_0) = 3/4 \pi$
(this is useful to simplify the notation), in this equation we introduce
$\mu_E = 2 l$ and $\mu_{E^1} =2 n$ as in the previous equation
\begin{eqnarray}
&& (3/4 \pi) \Big(\sqrt{|m - 1|} \, \varphi_{m - 1} -
    \sqrt{|m + 1|} \, \varphi_{m + 1} 
     - i  \, \sqrt{|m|} \, \varphi_{m} \Big) \phi_{2 l, 2 n} \nonumber\\ 
&& - \, \sqrt{3 | l |/ \pi} \, \varphi_{m} \big(\phi_{2 l, 2 n-1}  - \phi_{2 l, 2 n+1} \big) = 0.
\label{Constr1ml}
\end{eqnarray}

\begin{enumerate}

\item For $l = 0$ we obtain a relation to express $\varphi_{m }$ in terms of 
$\varphi_{m +1}$ and $\varphi_{m-1 }$. 
In particular introducing the initial conditions $\varphi_{m_0 -1}$ and $\varphi_{m_0 +1}$
(that are the same initial condition introduced in (\ref{EqDisInm}))
we can obtain $\varphi_{m_0}$ from equation 
\begin{eqnarray}
 \sqrt{|m_0 - 1|} \, \varphi_{m_0 - 1} -
    \sqrt{|m_0 + 1|} \, \varphi_{m_0 + 1} 
     - i  \, \sqrt{|m_0|} \, \varphi_{m_0}  = 0.
\label{Constr1ml=0}
\end{eqnarray}
At this point we have the even component $\varphi_{m_0}$ 
useful to solve at the same time equation (\ref{EqDisInm}) and equation 
(\ref{Constr1ml=0}) (see point 1.).

We observe that if we introduce $m = 0$ (and obviously $l=0$) in equation (\ref{Constr1ml})
we obtain $\varphi_{-1} = \varphi_{1}$, $\varphi_0$ decouples from the difference equation;
this means that not all the initial boundary conditions 
$\varphi_{m_0-5}, \, \varphi_{m_0-3}, \, \varphi_{m_0-1}, \, 
\, \varphi_{m_0+1}$, $\varphi_{m_0+3}$
are independent; there is one consistency relation between the five initial conditions.

\item For $l \neq 0$ we insert the solution $\varphi_{m}$ of
equation (\ref{EqDisInm}) in (\ref{Constr1ml}) and we obtain a second order 
difference equation for $\phi_{2 l, 2n}$ in the variable $n$.
Using the constraint (\ref{LastCon}) for $\sin^2(\chi_0) = 3/4 \pi$ we can 
evaluate equation (\ref{Constr1ml}) on 
$m = \mu_E = 2 l$ to obtain 
\begin{eqnarray}
 && (3/4 \pi) \left(\sqrt{|m - 1|} \, \varphi_{m - 1} -
    \sqrt{|m + 1|} \, \varphi_{m + 1} 
     - i  \, \sqrt{|m|} \, \varphi_{m} \right)\Big|_{m= 2 l} 
      \phi_{2 l, 2 n} \nonumber \\
&& \hspace{0cm} - \, \sqrt{3 | l |/ \pi} \, \varphi_{m}\Big|_{m= 2 l} 
\big(\phi_{2 l, 2n-1}  - \phi_{2 l, 2n+1} \big) = 0.
\label{Constr1mlmu}
\end{eqnarray}

Introducing the initial condition
$\phi_{2l, 2n_0-1}$ $\forall l$ and $\phi_{2l, 2n_0}$ $\forall l$
(this second condition can 
be obtained from equation (\ref{solutionln})),
we can calculate $\phi_{2l, 2n_0+1}$. This observation implies that
in the variable ``n", 
from equation (\ref{solutionln}) we can calculate the even components 
$\phi_{2 l, 2n}$ and from  (\ref{Constr1mlmu}) 
we calculate the odd components $\phi_{2l, 2n+1}$
of the wave function .

\end{enumerate} 
 
 \end{enumerate}
 
 We conclude the section summarizing the results about physical states. 
 Using the news notations we can write the physical states in the following way
 \begin{eqnarray}
 \langle \psi | = \sum_{m} \varphi_{m} \langle m | \otimes   \sum_{l, n} \phi_{2l, n} 
\langle 2l, n|,
\label{TensorStateFinal}
 \end{eqnarray}
where $\varphi_m$ is calculated using the two difference equations (\ref{EqDisInm})
and (\ref{Constr1ml=0}), 
and $\phi_{2l, n}$ can be obtained from equations (\ref{solutionln}) and (\ref{Constr1mlmu}).

\section*{Conclusions}

In this paper we have studied the gravitational collapse in Ashtekar variables following
the paper \cite{NS}. 
We have studied the gravitational collapse when all dust matter
has crossed the event horizon. In this particular region the metric is 
homogeneous and we have applied the technology developed in `` loop quantum 
cosmology" and in `` loop quantum black hole" papers \cite{MAT}, \cite{BojRP} and \cite{work4}.
We have divided the space-time region inside the event horizon in two parts, 
the firs one where the matter 
is localized and the other one part outside the mater. The space-time metric
inside the matter is of Friedmann-Robertson-Walker type and the metric 
outside the matter is of Kantowski-Sachs type with spatial topology  $\mathbf{R} \times \mathbf{S}^2$.

The quantization procedure is induced by  full ``loop quantum gravity".
We have introduced homogeneous holonomies 
and we have expressed the Hamiltoninan constraint in terms of such holonomies
in all the region inside the event horizon. 
The main result is that the Hamiltonian constraint gives a regular difference
equation for the coefficients of the physical states which are defined in the 
dual space of the dense subspace of the kinematical Hilbert space. 
We can summarize this result recalling that the quantum Einstein's equations
\begin{eqnarray}
&& ( \psi| \hat{H}_{E}(h, V) = 0,  \nonumber \\
&& ( \psi| (\hat{H}^{RW}_{E}(h^{RW},V^{RW}) + \hat{H}_M) = 0,
\end{eqnarray} 
and the boundary conditions are regular in $r=0$ where the classical 
singularity is localized ($h, V$ and $h^{RW}, V^{RW}$ are respectively the holonomy and the 
volume inside and outside the matter).

An important consequence of the quantization is that, unlike the classical 
evolution, the quantum evolution does not stop at the classical singularity
and the ``other side" of the singularity corresponds with a new domain where
the triad reverses its orientation. In this simply model we have solved 
the quantum Einstein equations obtaining recursive equations for  the coefficients 
$\psi_{\mu, \mu_E, \mu_{E^1}}$
of the physical states 
\begin{eqnarray}
\langle \psi | = \sum_{\mu, \mu_E, \mu_{E^1}} \psi_{\mu, \mu_E, \mu_{E^1}} 
\langle \mu, \mu_E, \mu_{E^1} |,
\label{BHstateConcl}
\end{eqnarray}
We can interpret $\psi_{\mu, \mu_E, \mu_{E^1}}$ as the wave function of the 
gravitational collapse, after the matter has crossed the event horizon.
We have also showed the consistency of the four difference equations.
In a future paper we will study the difference equations (\ref{solution}), (\ref{EqDisIn}) and (\ref{Constr1}) to obtain a numerical solution of those equations for a particular
value of the boundary condition on the wave function.

We hope that the analysis performed here will shed light on 
the problem of the ``information loss" in the process of black 
hole formation and evaporation. See in particular \cite{AB} for a possible
physical interpretation of the black hole information loss problem.

\section*{Acknowledgements}
We are grateful to Alfio Bonanno, Roberto Balbinot, Carlo Rovelli and  Eugenio Bianchi 
for many important,
instructive and clarifying discussions.

\end{document}